\documentclass{elsart}

\usepackage{epsfig}

\begin{document}

\def\npdg{$\overrightarrow{n} + p \rightarrow d + \gamma$}

\begin{frontmatter}

\title{
A measurement of parity-violating
gamma-ray asymmetries in polarized cold neutron capture
on $^{35}$Cl, $^{113}$Cd, and $^{139}$La  
}

\author[LANL]{G.S.~Mitchell\corauthref{GSM}},
 \corauth[GSM]{Corresponding author. Tel.: + 1-505-665-8484.}
 \ead{gmitchell@lanl.gov}
\author[IU]{C.S.~Blessinger\thanksref{cur1}},
\author[LANL]{J.D.~Bowman},
\author[UMichigan]{T.E.~Chupp},
\author[UMichigan]{K.P.~Coulter},
\author[IU]{M.~Gericke\thanksref{cur1}},
\author[Hamilton]{G.L.~Jones},
\author[UNH]{M.B.~Leuschner\thanksref{cur2}},
\author[IU]{H.~Nann},
\author[UManitoba]{S.A.~Page},
\author[LANL]{S.I.~Penttil{\"a}},
\author[Dayton]{T.B.~Smith},
\author[IU]{W.M.~Snow}, 
\author[LANL]{W.S.~Wilburn}

\address[Dayton]{University of Dayton, Dayton, OH 45469, USA}
\address[Hamilton]{Hamilton College, Clinton, NY 13323, USA}
\address[IU]{Indiana University, Bloomington, Indiana 47405, USA}
\address[LANL]{Los Alamos National Laboratory, Los Alamos, New Mexico 87545, USA}
\address[UManitoba]{University of Manitoba, Winnipeg, Manitoba R3T 2N2, Canada}
\address[UMichigan]{University of Michigan, Ann Arbor, Michigan 48104, USA}
\address[UNH]{University of New Hampshire, Durham, New Hampshire 03824, USA}

\thanks[cur1]{Current address: Los Alamos National Laboratory, 
 Los Alamos, New Mexico 87545, USA.}
\thanks[cur2]{Current address: Indiana University, 
 Bloomington, Indiana 47405, USA.}

\begin{abstract}
An apparatus for measuring parity-violating asymmetries
in gamma-ray emission following polarized cold neutron capture was 
constructed as a 1/10th scale test of the design for the forthcoming
\npdg\ experiment at LANSCE.  
The elements of the polarized neutron 
beam, including a polarized $^3$He neutron spin filter and a radio 
frequency neutron spin rotator, are described.  
Using CsI(Tl) detectors and photodiode current 
mode readout, measurements were made of asymmetries in gamma-ray 
emission following neutron capture on 
$^{35}$Cl, $^{113}$Cd, and $^{139}$La targets.
Upper limits on the parity-allowed asymmetry 
$ s_n \cdot (k_{\gamma} \times k_n)$ 
were set at the level of $7 \times 10^{-6}$ for all three targets.
Parity-violating asymmetries $ s_n \cdot k_{\gamma} $ 
were observed in $^{35}$Cl, $ A_\gamma = (-29.1 \pm 6.7) \times 10^{-6} $, 
and $^{139}$La, $ A_\gamma = (-15.5 \pm 7.1) \times 10^{-6} $,
values consistent with previous measurements.
\end{abstract}

\end{frontmatter}

\section{Introduction}
\label{}

The NPDGamma experiment \cite{Snow1,Snow2} is under construction 
at the Los Alamos Neutron Science Center (LANSCE) at Los Alamos 
National Laboratory. This includes construction of Flight Path 12 
at the Manuel Lujan Jr.\ Neutron Scattering Center at LANSCE, 
which will be a pulsed cold neutron beamline dedicated to 
fundamental physics.  The goal of the experiment is to measure the 
parity-violating directional gamma-ray
asymmetry $A_\gamma$ in the reaction \npdg\ to an 
accuracy of $5\times10^{-9}$, which is approximately 10\% of its 
predicted value~\cite{DDH}.  This measurement will provide a 
theoretically clean determination of the weak pion-nucleon 
coupling and resolve the long-standing controversy over its 
value~\cite{AH,HH,Wilburn}.  The experiment will consist of a pulsed, cold 
neutron beam, transversely polarized by transmission 
through polarized $^3$He, incident on a liquid para-hydrogen
target.  The 2.2 MeV gamma-rays from the capture reaction will be 
detected by an array of CsI(Tl) scintillators coupled to 
vacuum photodiodes and operated in current mode.  In Fall 2000,
an engineering run was completed using prototypes of all major 
components to measure parity-violating asymmetries in neutron 
capture on several nuclei.  Accuracies of order 
$7\times10^{-6}$, limited by counting statistics, were obtained 
after several hours of running using Cl, Cd, and La 
capture targets.  This paper will discuss the results of this 
engineering run and its implications for the design of the
NPDGamma experiment.

Parity violation permits a term in the differential cross section for
the ($n$,$\gamma$) reaction 
proportional to $ s_n \cdot k_{\gamma} $, where $s_n$ is the neutron 
spin direction and $k_{\gamma}$ is the photon momentum vector. 
A constant, $A_\gamma$, measures the size of this term 
in the differential cross section for gamma-ray emission. 
The cross section is proportional 
to $1+A_\gamma \cos\, \theta$, where $\theta$ is the angle between the 
neutron polarization and photon momentum. The parity violation
arises due to weak interactions inside and between the nucleons in the
nucleus, which introduces new opposite-parity components
into the initial and final states and allows mixing and interference 
between electromagnetic transitions from opposite parity
states~\cite{AH}.  For example, in the \npdg\ reaction 
weak effects allow a small amount of E1 transition to 
interfere with the primary M1 amplitude. In systems with 
$Z>1$ the interference is typically much more complicated,
involving many states and many transitions.
Parity-allowed asymmetries in the differential cross section with
nontrivial angular distributions such as 
$ s_n \cdot (k_{\gamma} \times k_n)$, where $k_n$ is the neutron 
momentum vector, are also possible~\cite{Csoto}. 
A general analysis of the various angular and polarization 
correlations in ($n$,$\gamma$) reactions
is given in~\cite{Flambaum}.

The $ s_n \cdot k_{\gamma} $ correlation has been observed 
in $^{35}$Cl and $^{139}$La in previous experiments \cite{Vesna,Avenier}. 
While parity violation is observed in neutron capture on $^{113}$Cd in
$p$-wave resonances at epithermal neutron energies, for cold neutron
capture the process is dominated by a strong $s$-wave resonance and no 
parity violation is expected.  The origin of the parity-violating
effect in $^{139}$La is known to be due to
mixing with a narrow $p$-wave resonance at 0.734 eV. 
The huge ($\sim 10$\%) parity-odd effects at resonance in
this~\cite{Yuan} and many other heavy nuclei~\cite{TRIPLE} are now 
understood in terms of two mechanisms:
dynamical enhancement, which comes from the close spacing of two
levels of opposite parity in the compound resonance regime; and 
kinematic enhancement, which is due to the difference in widths 
of the $s$ and $p$ resonances involved in the interference.
The size of the effect for cold neutron energies below the $^{139}$La 
resonance is as expected from the tail of this resonance.    
The origin of the parity-violating effect in $^{35}$Cl is thought 
to be due to the mixing of two opposite-parity levels, a 
$J^{\pi}=2^{-}$ $p$-wave level at 398 eV and a 
$J^{\pi}=2^{+}$ subthreshold $s$-wave
resonance at -130 eV. 
The presence of a $p$ wave in the intermediate state in combination 
with final state effects in the reaction can also give rise to the 
parity-allowed $s_n \cdot (k_{\gamma} \times k_n)$ 
correlation. It was therefore possible that a significant parity-allowed 
asymmetry in gamma-ray emission following polarized cold neutron capture 
in $^{35}$Cl or $^{139}$La might exist.

There are several motivations to 
measure parity-violating and parity-allowed asymmetries on Cl, La, 
and Cd targets in preparation for the \npdg\ experiment. 
Foremost, the measurement of the parity-violating correlations 
can be used to test a 1/10th scale version of the planned apparatus 
at the few parts per million  
level, with the Cd target as a null test.  In addition, 
the discovery of a large parity-allowed $ s_n \cdot (k_{\gamma} \times k_n)$ 
asymmetry would be useful for a detector alignment scheme for the 
NPDGamma experiment.  Knowledge of the detector element angles to
a precision of 20~mrad with respect to the neutron spin direction 
(determined by a magnetic holding field) is 
required in order to suppress systematic effects associated with
parity-allowed neutron spin-correlated gamma-ray signals 
leaking into the orthogonal direction associated with the parity
violation signal.  Finally, the measurement also provided an opportunity to
check calculations of the neutron moderator 
brightness and to measure relative
intensity fluctuations in the neutron beam 
and limit this potential source of extra noise into the \npdg\ measurement.

\section{Description of Setup}

\begin{figure}[tbp]
\begin{center}
\mbox{\epsfig{file=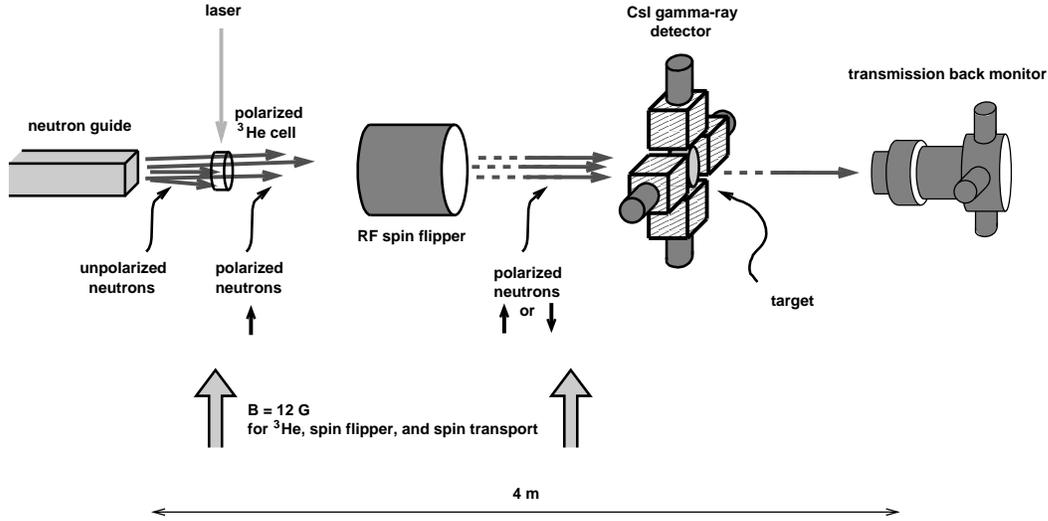,width=5.5in,angle=0}}
\end{center}
\caption{Experimental setup for asymmetry measurements on nuclear targets.  
Various elements of collimation and shielding and the
large coils providing the vertical 12~G field over the 
entire apparatus are not shown.
The magnetic field covered the region from the end of the guide
to the target.}
\label{fig:setup}
\end{figure}

This section describes the apparatus used in Fall 2000 to 
measure directional asymmetries in gamma-ray emission following
polarized cold neutron capture on nuclear targets.  A schematic
picture of the setup is shown in Fig.~\ref{fig:setup}.

\subsection{Pulsed Cold Neutron Beam}

The LANSCE linear accelerator provides 800 MeV protons to a
proton storage ring, which after compression of the beam 
delivers 250~ns wide (at the base) pulses at 20~Hz to a split tungsten
spallation target.  The downstream target is surrounded by several
moderators, including a partially coupled cold hydrogen moderator
viewed by Flight Path 11A (FP11A).  
The performance of the moderator is modeled in \cite{Ferguson}.  
The neutron flux is directed through FP11A by a 
$^{58}$Ni-coated $6\times6$ cm$^2$
guide which begins 1~m from the moderator and is 18.5~m long.
On the scale of 50~ms between pulses, the proton pulse width is 
very short and thus has an insignificant effect on the width
of the neutron pulse from the moderator.  
An experiment located $\sim 20$~m from the source will 
see a time of flight spectrum of cold neutrons: for the typical
energies of interest (1--100~meV, 1--9~\AA) 
neutrons will arrive 4.5--45~ms after the proton
pulse.  Knowledge of time of flight and the flight path 
length determines the neutron energy.

\subsection{Measurement of Cold Neutron Flux}

An absolute measurement of the cold neutron flux was made
by tightly collimating the beam with a 4~mm diameter Gd foil
collimator.  The collimator was placed in a shielding assembly 
15~cm from the end of the guide.
A 4.1~mm diameter $^6$Li-loaded glass scintillator was mounted
on a photomultiplier tube which was itself mounted in an x-y
positioner. This detector was located 2.65~m from the end of
the guide.  The collimation reduced the 
neutron rate enough to permit normal pulse counting techniques (to less than
60 kHz for energies less than 15~meV). With incident proton
current of 90~$\mu$A,
the peak neutron flux out of the end of the uncollimated FP11A guide
was $1.4\,\times\,10^6$ neutrons/ms per pulse at 8~meV, and the 
total flux from 2.5 to 40~meV was $2.5\,\times\,10^7$
neutrons per pulse.  (Over the same energy range, 
including collimation to 5~cm diameter and transmission
through the polarizer, there were $2.7\,\times\,10^6$ neutrons/pulse 
incident on the target for the asymmetry measurements discussed below.)
Using the geometry of the flight path and the collimation to 
convert from the measured rates, the result was a measured peak 
moderator brightness 
of $8.6 \times 10^7$ neutrons/cm$^2$/s/sr/meV/$\mu$A at 
4~meV, which is  20\%  lower
than predicted by~\cite{Ferguson}.  
While the new FP12 is predicted to have a 50\% larger brightness 
than FP11A~\cite{FP12},
scaling by the measured FP11A flux indicates that FP12
will only be sufficient for the full NPDGamma experiment to make a
raw asymmetry measurement of $1.4 \times 10^{-4}$ per neutron pulse.
This corresponds to a physics asymmetry measurement of 
$3.7 \times 10^{-4}$ per pulse.  
Thus in the planned run time of three six month run cycles, 
where a run cycle consists of 2500 hours of 120~$\mu$A proton
beam, the predicted sensitivity of the experiment 
to $A_\gamma$ is at the level of $1.6 \times 10^{-8}$ (30\% of
the DDH prediction~\cite{DDH}).

\subsection{Intensity Fluctuations}

Several possible processes, such as density fluctuations in the 
moderator or position fluctuation of the
proton beam striking the tungsten spallation target, can produce 
pulse-to-pulse position fluctuations of the neutron beam which will 
induce noise in the asymmetry measurement that is greater than the 
counting statistics of the experiment. For the NPDGamma experiment, 
noise sources must be negligible compared to neutron 
counting statistics.  A simple Monte Carlo study of 
intensity-induced position fluctuations of the neutron 
beam showed that relative intensity fluctuations with a 
variance of $\sigma^2 \approx 10^{-4}$ lead to noise in the asymmetry 
measurement equal to neutron counting statistics. Three sets of 
measurements were made of correlations between 
a proton current toroid monitor and a $^3$He ion chamber 
neutron flux monitor.  The neutron intensity fluctuations 
for constant incident proton current were 
$\sigma^2 \approx 10^{-5}$ for each set of measurements, 
an order of magnitude better than the requirement.
These measurements of the intensity fluctuations of the beam for 
constant proton current incident on the spallation target 
show that density fluctuations or bubbles in the 
cold hydrogen moderator were not a problem for the asymmetry
measurements in this engineering run, and 
will not be a problem for NPDGamma.

\subsection{$^3$He Spin Filter}

Cold neutron beams can be polarized in several ways, but
the best technology for NPDGamma is a $^3$He spin 
filter \cite{Jones}.  $^3$He spin filters are compact, possess 
a large phase space acceptance,  and produce a negligible fraction 
of capture gamma-ray background.  In contrast to a polarizing supermirror, 
a $^3$He spin filter does not require strong
magnetic fields or produce field gradients. This is important for the 
control of systematic errors in the NPDGamma measurement. The 
direction of the neutron beam motion due to Stern-Gerlach steering 
in a magnetic field gradient is correlated with the neutron spin 
direction and can therefore lead to a false asymmetry due to the 
solid angle change of the capture gamma-ray distribution in the target 
as seen by the gamma-ray detectors.
For this engineering run a system with a double-chambered 
glass cell was used to contain the $^3$He.
The cell also contained a small amount of Rb and N$_2$.
The $^3$He was polarized by spin-exchange optical pumping.
The polarization of the $^3$He proceeds by optically
pumping the 795~nm $5s - 5p$ transition for the Rb valence
electron, which transfers its spin to the $^3$He
nucleus by a hyperfine interaction upon collision.  
The cross section
for cold neutron absorption by $^3$He is essentially zero
for parallel neutron and nuclear spins, but due to a
large spin zero resonance in $^4$He, there is a large absorption 
cross section for anti-parallel spins.  Thus a polarized
volume of $^3$He can filter out one spin state of an unpolarized
neutron beam and produce large neutron polarizations.
The thickness of the spin filter can be optimized for 
polarization versus transmission.

A 12~G holding field was used to 
provide a uniform field parallel to the polarization direction
and to prevent depolarization of the $^3$He by diffusion in 
transverse magnetic field gradients.  Circularly polarized 
795~nm laser light for electronic polarization of the Rb was 
provided by two 15~W laser diode arrays.  One chamber of 
the cell, the pumping chamber, was in a small oven which 
was kept at 175~$^\circ$C to control Rb vapor density, and this
part was illuminated by the diode lasers;  the other
part of the cell, the polarizer, was the volume placed in the 
neutron beam and used to filter out the undesired polarization
state.  The polarizer chamber as seen by the neutron
beam was circular, 7~cm in diameter,
and 1.6~cm thick.  Both upstream and downstream surfaces
were concave with respect to the beam direction, in order to 
provide a uniform cell length to the beam while maintaining 
curved surfaces to prevent cell explosion.  The polarizer 
chamber was connected by a 1~cm long glass feedthrough 
tube to the roughly spherical and 6~cm diameter pumping chamber.

The absolute neutron polarization was measured using
a supermirror~\cite{Schaerpf} and an ion chamber
back monitor (discussed below), and the 
$^3$He polarization direction was reversed using an NMR adiabatic fast 
passage spin flip. Relative measurements of the $^3$He polarization 
were performed during data taking by an NMR system. 
Since in the test run it was discovered that the stray magnetic fields
from the supermirror polarization analyzer interfered with the
operation of the neutron spin rotator, the \npdg\ experiment
will use a polarized $^3$He neutron polarization analyzer instead. 

\begin{figure}[tbp]
\begin{center}
\mbox{\epsfig{file=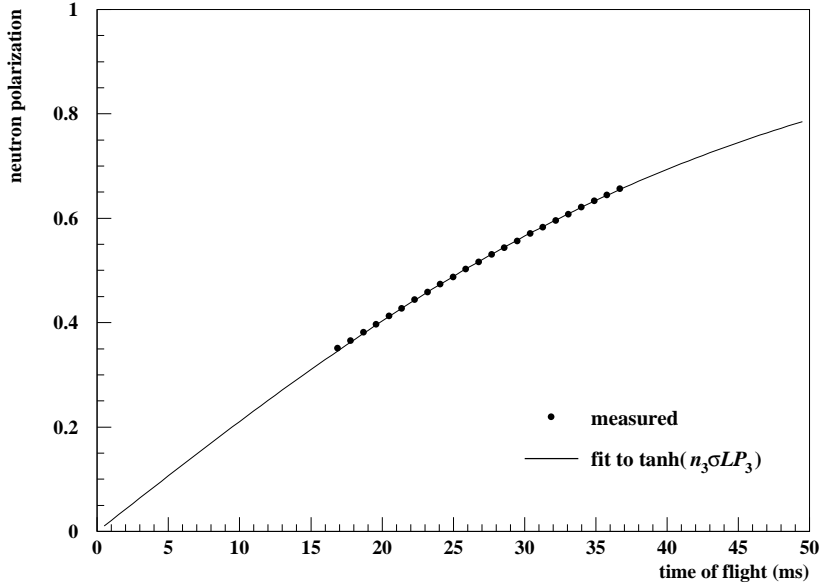,width=5.2in,angle=0}}
\end{center}
\caption{Neutron polarization. Measured points were obtained using
a supermirror polarization analyzer
and an ion chamber to compare neutron transmission of 
opposite $^3$He polarization directions.  The curve is a fit to 
$P_n = \tanh (n_3 \sigma L P_3)$, with $P_3$,
the $^3$He polarization, as a fitted parameter, 
using $n_3 L = 6.0\ {\rm atm\ cm}$  for the thickness 
of the $^3$He cell. 
The flight path length to the $^3$He cell was 21.08~m.
The flight path length to the back monitor 
was 23~m, which is used here to relate
the time of flight to neutron energy and $\sigma$,
the $^3$He cross section.  The uncertainty in the absolute scale
of the polarization is 2\%.}
\label{fig:3He}
\end{figure}

Measured neutron polarization varied from $P = 0.3$ to $0.7$
for neutron energies of 10~meV to 2~meV, shown in Fig.~\ref{fig:3He}.
The neutron polarization is given by $P_n = \tanh (n_3 \sigma L P_3)$
where $n_3$ is the number density of $^3$He, 
$\sigma = (26.8 / \sqrt{E} )$~kb,
$E$ is neutron energy in meV, $L$ is the length of the polarizer
volume, and $P_3$ is the $^3$He polarization.
At typical operating temperature, 
the 6.0~atm~cm polarizer section of this cell
had $^3$He polarization of $P_3 = 0.265$. The relative error
on the neutron polarization is 2\%, a combination of three
roughly equal sources of error from knowledge of the supermirror 
analyzing power, 
absolute accuracy of the NMR system, and statistical spread 
of individual NMR measurements.

\subsection{Spin Flipper}

To isolate a small parity-violating asymmetry, a common technique
is to reverse the polarization of the incident beam and observe a 
correlated change in the direction of emission of the reaction
products.  For NPDGamma and for this test run, the neutron spins 
are flipped on a 20~Hz pulse-by-pulse basis with a radio
frequency spin rotator, or spin flipper (RFSF).  
The RFSF is a shielded solenoid,
30~cm in diameter and 30~cm long, with the aluminum shielding 
canister a total of 10~cm larger in each dimension.
The windows encountered by the neutron beam are 2 mm thick.
The RFSF is similar to that employed in \cite{Alvarez}
with two modifications: the aluminum shielding, and the amplitude
of the RF field is ramped down with an inverse time dependence 
every pulse to match the time the neutrons spend traversing 
the length of the coil. 

The RFSF operates according to the well-known principles of NMR. 
In the presence of a homogeneous DC magnetic field 
and an oscillating magnetic field in a perpendicular direction, 
the neutron spin will precess \cite{Abragam}.
For a DC field of $B_0 \hat{y}$ and an RF field given by 
$\pol{B_{rf}} = B_1 \cos \left( \omega t \right) \hat{z}$,
transforming into a frame which rotates 
(at the same frequency $\omega$ as the oscillating field)
about the DC magnetic field direction, the effective field is
\begin{equation}
\pol{B} = \left( B_0 - \frac{\omega}{2 \pi \gamma} \right) \hat{y} 
+ \frac{B_1}{2} \hat{z}
\end{equation}
where 
$\gamma = 2 \mu / h = -2.916 \times 10^7$~Hz/T is the neutron
gyromagnetic ratio.  
For $\omega = 2 \pi \gamma B_0$, a neutron initially polarized 
along $\hat{y}$ will precess about the effective field direction
$\hat{z}$.  If the RF field is applied for a time equal to
$\tau = {1}/({\gamma B_1}) $
then the neutron spin direction will be reversed relative to 
$\hat{y}$.

The RFSF limits the region of the RF field 
along the neutron beam line by placing the RF coil inside of 
a cylindrical aluminum shield.  The RF coil is a solenoid 
with its axis aligned along the neutron beam direction
$\hat{z}$.  For neutrons from a spallation source and 
a RF interaction region of fixed length $L$, the interaction time 
$T$ is determined by the neutron velocity $v(t)$.  The time 
$T$ will increase with increasing time of flight $t$ and obey 
the relationship 
\begin{equation}
T = \frac{L}{v(t)} = \frac{L t}{d}, 
\end{equation}
where $L$ is the length of the RFSF, and $d$ is the distance 
from the spallation neutron source to the RFSF.  To produce a 
spin flip ($T = \tau$), the RF amplitude must obey the relationship 
\begin{equation}
B_1(t) = \frac{d}{\gamma L t}.
\end{equation}
In practice, the RF amplitude is not constant inside the 
RFSF since the normal component of the field must go to zero at 
the endcaps.  
Instead, it is the integral of the RF magnetic field 
amplitude along the length of the RFSF which
must vary inversely with time to meet the spin flip condition 
for all of the neutron energies. 

This type of spin rotator is an ideal choice for the
NPDGamma experiment for a number of reasons. 
First, the spin flip can be performed on a pulse-by-pulse basis
by simply turning the RF field on and off. 
Second, since one can contain the RF magnetic fields in metallic
windows that are transparent to neutrons and the gain of the CsI
detectors with vacuum photodiodes is relatively insensitive to DC and AC
magnetic fields, couplings of the RF signal into the data stream
which could lead to false asymmetries are greatly minimized. 
Third, no extra static magnetic fields or field gradients beyond those
required for neutron spin transport are required.
Fourth, the RFSF provides efficient pulse-to-pulse control 
of neutron polarization state for a wide range of energies.
Fifth, and finally, this type of spin flipper does not change the neutron
beam phase space or kinetic energy.  Although a neutron will absorb
or emit energy due to the external RF magnetic field, on resonance this
change in energy is balanced by the change in potential energy of
the neutron in the static magnetic field and the neutron kinetic
energy does not change \cite{Gol94}. Therefore a host of possible
systematic effects associated with a spin-dependent neutron
energy spectrum of the beam are absent.

The RFSF field amplitude and frequency were tuned using
the polarized neutron beam produced by the $^3$He system and
a supermirror analyzer \cite{Schaerpf} to analyze the flipped
pulses. The supermirror analyzer uses the spin dependence of the
reflection probability of polarized neutrons from the neutron
optical potential of magnetized mirrors to analyze the
polarization with an analyzing power approaching 100\% for cold
neutrons.  

The $^3$He/H$_2$ ion chamber back monitor 
was placed downstream of the supermirror
and the difference it recorded between RFSF on and off pulses 
yielded the RFSF efficiency.  The RFSF was typically driven
at 34~kHz in the holding field of 12~G, a relationship
determined by the neutron gyromagnetic ratio.
For a 2.5~cm collimated beam, the RFSF efficiency, defined as the
absolute ratio of transmitted (flipped) neutron 
polarization to incident polarization,
was measured to be $\epsilon = 0.98\pm0.02$.
This efficiency is sufficiently large and well-known for NPDGamma.

\subsection{Beam Monitors}

Two current-mode beam monitors were used in this test run:
a front monitor with $^3$He and $^4$He, and a back
monitor with $^3$He and H$_2$. The design of the front monitor
allows it to employ an analog subtraction between the current
signal of the $^3$He-filled front chamber  (which responds to
neutrons and gamma-rays) and the current signal from the
$^4$He-filled rear chamber (which responds only
to gamma-rays since the neutron absorption cross section of $^4$He is
zero) to produce a signal which is dominated by neutron capture
in the monitor~\cite{Szymanski94}. As discussed above, 
the front monitor was used to measure fluctuations in 
beam intensity. It was then removed from the beamline for the
measurements of gamma-ray asymmetries.

The back monitor, which is a $^3$He-filled segmented ion chamber
similar in design to~\cite{Penn}, was 
used to study the $^3$He spin filter system and the RFSF. It
consists of a metal cylinder coaxial with the neutron beam with 
longitudinal segments bounded alternately by high voltage
electrodes and signal collection plate electrodes. 
The chamber housed ten 7.6~cm diameter, 0.76~mm thick Al 
collection plates spaced evenly by 2.5~cm increments along 
the beam axis, and ten electrodes held at 3~kV. 
The voltage as a function of depth along the beam axis
makes a zigzag pattern, with ions on either side of a high
voltage plate accelerated to opposite signal plates. 
The detector is therefore effectively segmented into separate
regions of sensitivity along the beam axis. This segmentation
along with the known neutron absorption and scattering cross
sections implies that the neutron signal changes in a known way as
the neutron energy spectrum hardens as it passes into the
chamber. A gas mixture of 0.5~atm $^3$He and 3~atm
H$_2$ was chosen based on simulations of the ion chamber
response. H$_2$ gas was chosen as the fill gas due to its low
sensitivity to gamma-rays and the order-of-magnitude greater mobility
of ions in  H$_2$ gas relative to other gases. This choice led
to a collection time for the ions of about 100 microseconds.  The
$^3$He gas pressure and overall length of the chamber was chosen
to produce an essentially unit neutron detection efficiency from
1--100 meV. If the detector
efficiency is flat and the ion chamber signal is dominated by
neutron capture, the observed current is proportional to
neutron flux. 

The depth dependence of the signal was consistent
with expectations based on the assumption that the current mode
signal was due to neutron capture. The sensitivity of the ion
chamber to gamma-rays was measured by converting the entire neutron
beam into 0.5 MeV gamma-rays directly in front of the chamber by
absorption by the $^{10}$B in B$_4$C. This test demonstrated 
that less than 0.3\% of
the signal was due to gamma-rays in the neutron beam. The neutron
current measured by the ion chamber shows discontinuities which
correspond to the presence of Bragg edges from the aluminum
material in the beam. These discontinuities correspond to neutron
wavelengths $\lambda=2d$ below which diffraction from those
lattice planes separated by $d$ are forbidden. The relative time of flight
positions of these Bragg edges in combination with the known
source-chamber distance and the known $d$ spacings of aluminum are
used to calibrate the time of flight spectrum in terms of neutron energy.
Materials of the chamber were chosen to be high voltage compatible and
bakeable to remove electronegative impurities and improve the ion
drift velocity. The signals from the collection plates were read
out to current to voltage amplifiers similar to those used for
the CsI detectors with a 100~M$\Omega$ gain resistor.

\subsection{Nuclear Targets}

For parity-violating and parity-allowed asymmetry measurements, 
three targets were
used.  As a chlorine target, an aluminum canister was
filled with CCl$_4$ (natural Cl).  The CCl$_4$ thickness 
seen by the neutron beam was 3.8~cm.  As a lanthanum target, 
a natural La (99.91\% $^{139}$La) metal cylinder 10.2~cm in diameter
and 2.5~cm thick was housed in a 1~mm thick stainless
steel can.  As a cadmium target, a 0.76~mm thick piece
of natural Cd metal was cut into a 10.2~cm diameter circle.
To eliminate neutron backgrounds and reduce activation of
the CsI(Tl), 
the targets were in turn mounted in a cylinder of $^6$Li-doped
plastic with inner diameter 10.2~cm and outer diameter 15.2~cm.
The neutron beam was collimated to 5~cm diameter and 
no neutrons were directly incident on the plastic cylinder.
The targets were located at a flight path length of 21.9~m.

\subsection{Gamma-ray Detectors}

To detect the gamma-rays emitted following neutron capture,
an array of four CsI(Tl) crystals was symmetrically mounted
around the target volume.  The crystals
were housed in Al canisters 
with external dimensions $15 \times 16.5 \times 16.5$
cm$^3$.  The closest faces of the detectors were 8.4~cm from
the central axis of the neutron beam, allowing each detector to 
cover a solid angle of $\sim 0.1 \times 4 \pi$~sr with respect
to the target location.   The crystals were
viewed with 70~mm vacuum photodiodes
(Hamamatsu R2046PT) biased at -90~V.  The photoelectron
yield of the crystals was measured to be $\sim 500$~p.e./MeV,
and typical photodiode currents were a few nA. This light output
was sufficient to make
the extra noise encountered in current mode measurement due to
fluctuations in the number of quanta produced per neutron capture
small compared to neutron counting statistics. 

The magnetic field sensitivity
of the photodiode signals to DC fields was measured to be less than 
$1 \times 10^{-4}$/G in a 10~G field directed perpendicular to the 
photodiode axis.
The second-order magnetic field sensitivity of the photodiode 
signals was measured to be less than $1 \times 10^{-5}$/G$^2$, 
in an oscillating field with amplitude of 10~G. 
For this engineering run and for the NPDGamma experiment,
the RFSF magnetic field is the field with the most
possibility for undesired experimental effects, as its presence or
absence is by definition correlated with the neutron spin state.
Since the RFSF magnetic field is not DC but
operates at $\sim$30~kHz, a fast time scale compared to
the recorded data, first order gain effects will cancel out,
but second order effects may not.  
The measured sensitivities 
combined with the fields and field variations expected in the 
NPDGamma experiment lead to negligible systematic effects.

\begin{figure}[tbp]
\begin{center}
\mbox{\epsfig{file=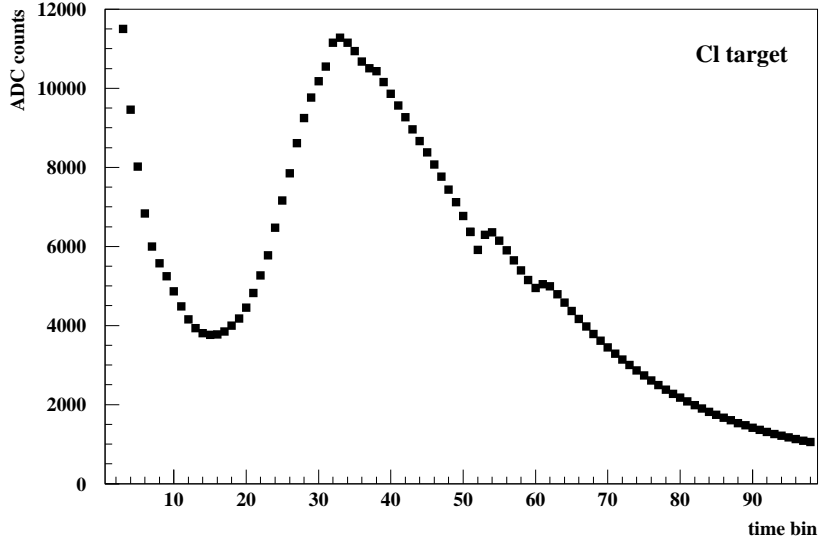,width=5.2in,angle=0}}
\end{center}
\caption{Average spectrum for a Cl target run.  The x axis of
time bins 0--100 corresponds to time of flight of 0--40~ms.  The
y axis of ADC counts is proportional to the voltage signal out of
the preamplifier, where one ADC count is 0.3~mV.
The dips in the spectrum at time bins greater than 35 are 
due to Bragg edges of materials such as aluminum windows
in the flight path.
}
\label{fig:waveform}
\end{figure}

The current signals from the photodiodes were read out 
into a low-noise solid state current to voltage 
preamplifier with a gain resistor of 10~M$\Omega$.
The preamplifier signals were sampled
at 25~kHz with 16 bit ADC's in a VME system, and the data compressed by
a factor of ten to be written as 100 samples, or time bins, over 40~ms
per pulse.  
With the neutron beam off, the r.m.s.\ noise per time bin
was measured to be of order 1~mV.  
Pedestal values for the detectors were typically 
100 ADC counts, or 30~mV.  An average gamma-ray intensity 
spectrum, indicative of the time structure of the
neutron flux, is shown in Fig.~\ref{fig:waveform}.

\section{Asymmetry Measurements}

Asymmetry measurements were made on three targets.  The neutron 
energy range analyzed in each case was 2.5 meV to 40 meV, 
corresponding to 8--32 ms time of flight, or 1.4--5.7 \AA.

Asymmetries were formed using matched eight-step
spin sequences ($\uparrow \downarrow \downarrow \uparrow 
\downarrow \uparrow \uparrow \downarrow$)
of consecutive pulses and a calculation of the geometric mean
asymmetry within the sequence. This 8-step sequence is chosen to cancel 
linear and quadratic time-dependent drifts in detector
efficiencies.  The raw experimental asymmetries are calculated as follows:
\begin{equation}
 A_{\rm raw} = {
\sqrt{U_\uparrow D_\downarrow \over U_\downarrow D_\uparrow} - 1
\over
\sqrt{U_\uparrow D_\downarrow \over U_\downarrow D_\uparrow} + 1
}
\end{equation}
where $U,D$ refer to the upper and lower detector signals, and the 
subscripts $\uparrow, \downarrow$ refer to the neutron
spin direction.  For these measurements the up-down direction was
the axis of the neutron polarization, so an asymmetry between the
$U$ and $D$ detectors is parity-violating.  An asymmetry between
the $L$ and $R$ detectors is parity-allowed.
The state $\uparrow$ corresponds to RFSF
on, $\downarrow$ to RFSF off.  The asymmetry $A_{\rm raw}$
was calculated for each valid eight-step spin sequence, in each 
case four steps added together to obtain $U_\uparrow$ and $D_\uparrow$
and the other four to obtain $U_\downarrow$ and $D_\downarrow$.  
For each sequence the raw asymmetries were calculated by time bin using the 
data from the 20th to 80th of the 
100 samples.  This range in neutron time of flight corresponds to the
peak of the neutron distribution and to a region of well-understood
neutron polarization and RFSF efficiency.
An identical procedure was used for the left and right detector pair.
A cut was made to eliminate sequences which had pulses 
with anomalous incident neutron fluxes.
If any of the eight steps had a detector sum ($U+D+L+R$) over all
time bins that differed by more than 1\% from the average of all
the sequences in the sum, the entire eight-step sequence was discarded. 
This cut removed less than 1\% of the data.  The final data
set for each target consisted of 6.5 to $8.4 \times 10^4$
eight-step sequences.

For each sequence, the measured asymmetries $A_{\rm raw}$ were 
corrected to physics asymmetries $A_\gamma$ in each time bin using
\begin{equation}
 A_\gamma = {(A_{\rm raw} - A_{\rm noise}) \over PRTG }
\end{equation}
where $A_{\rm noise}$ is a measured false asymmetry due to 
electronic noise pickup, 
$P$ is the neutron polarization,
$R$ is the RFSF efficiency, 
$T$ is a factor for neutron depolarization in the target prior to capture, and 
$G$ is a geometry factor for the average angle of the gamma-rays seen
by the detectors relative to the neutron polarization.
These factors are discussed in the following five paragraphs.
The factors $P$ and $T$ vary with neutron energy, while the
others are independent of the neutron energy (constant for
all time bins).  The data were combined in this way in order to
obtain the proper weighting of events at different energies.

The electronic noise asymmetry was measured before and after 
the neutron capture data were acquired.  With no neutrons
incident on the apparatus, the average of two one-hour runs was
$A_{\rm noise}^{UD} = (0.12 \pm 0.26) \times 10^{-6}$
and
$A_{\rm noise}^{LR} = (0.42 \pm 0.25) \times 10^{-6}$.
These values were extracted from the data using the same
method as for $A_{\rm raw}$.  Since these values are less
than two standard deviations from zero and are an order of
magnitude smaller than the statistical error, $A_{\rm noise}=0$
is used to obtain physics asymmetries from the raw asymmetries.

The absolute neutron polarization was measured using the supermirror
analyzer and back monitor neutron detector, 
and fit to a hyperbolic tangent function
to relate the pickup coil NMR signal to $P_3$.  NMR measurements
were made during data taking and the NMR signal amplitudes 
measured during data taking with each target
were used to provide a $P_3$ for the hyperbolic tangent expression
given earlier for neutron polarization versus energy. 
The convolution of the neutron polarization values and the
detector signals (versus energy) leads to an event weighted
average polarization of $P\approx 0.37$ for each of the targets.

Since the RFSF is on for half the neutron pulses,
the measured efficiency $\epsilon =0.98 \pm 0.02$ 
yields a factor of $R = (1+\epsilon)/2 = 0.99 \pm 0.01$
for converting to a physics asymmetry.

Neutron depolarization in the target, $T$, is accounted for
using cross section values for ($n,\gamma$), spin-coherent
scattering, and spin-incoherent scattering, taken from~\cite{Sears}.
A simple Monte Carlo was written to propagate neutrons through
the target material according to the cross sections (assuming a $1/v$
dependence of the capture cross section for $^{35}$Cl and $^{139}$La), 
and upon
each scattering event the probability of spin-flip scattering
was accounted for by taking the ratio of $\frac{2}{3}$ of the spin-incoherent 
scattering cross section to the total scattering cross section.  
Depolarization was determined by computing an average value
for $(-1)^n$ where $n$ was the number of spin-flip scatterings
prior to capture.
The convolution of the depolarization values with the detector signals 
leads to event weighted factors of  $T = 0.92, 1.00, 0.94$ 
for Cl, Cd, and La respectively.
For each target, it was assumed that isotopes other than the 
one of interest ($^{35}$Cl, $^{113}$Cd, and $^{139}$La)
contribute negligibly to the scattering, capture, or $A_\gamma$.

The geometrical acceptance of the detector crystals was modeled
using the computer code MCNP~\cite{MCNP}.  The result for average $\cos (\theta)$
for parity-violating (up-down) asymmetry and for average $\sin (\theta)$ for
the parity-allowed (left-right) asymmetry 
yielded factors of $G = 0.86, 0.86, 0.87$ for Cl, Cd, and La respectively.
The angle $\theta$ is that between the neutron polarization axis
and the gamma-ray momentum.  

\begin{figure}[tbp]
\begin{center}
\mbox{\epsfig{file=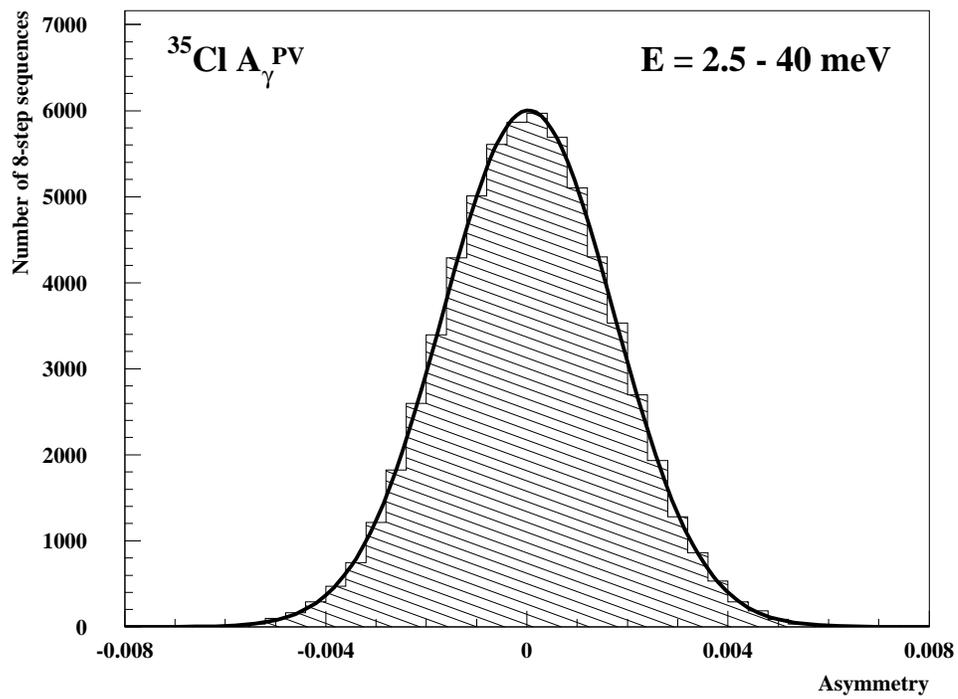,width=5.0in,angle=0}}
\end{center}
\caption{Histogram of asymmetry values from eight-step sequences
for the Cl target.  A fit to a Gaussian distribution is
shown. The mean and the error in the mean of the distribution
were calculated not from the fit but 
by simple averaging of the list of values.}
\label{fig:asymmetry}
\end{figure}

Following the above analysis, each sequence produced a result
for parity-violating $A^{\rm PV}_{\gamma}$ 
and parity-allowed $A^{\rm PA}_{\gamma}$,
corresponding to $s_n \cdot k_{\gamma}$ and 
 $ s_n \cdot (k_{\gamma} \times k_n)$, where 
$s_n$ is the neutron spin direction, $k_n$ is the neutron momentum vector,
and $k_{\gamma}$ is the photon momentum vector.
A histogram of the values of $^{35}$Cl
$A^{\rm PV}_{\gamma}$, with one entry per sequence, is
shown in Fig.~\ref{fig:asymmetry}.
The physics asymmetries for each target were
histogrammed and fit by minimizing $\chi^2$ compared to a Gaussian 
distribution to extract the mean value and the error in the mean.  
The results of the fitting
are consistent with the simple average of each data set.
For final results, the values for $A^{\rm PV}_{\gamma}$ 
and $A^{\rm PA}_{\gamma}$ are obtained from taking 
the list of sequence asymmetry values and calculating the
mean and error in the mean from the simple average and 
standard deviation of the data.
The signs of the asymmetries were carefully checked through
the data acquisition electronics and analysis code.  

Parity-violating asymmetries were detected (greater than two
standard deviations from zero) in neutron
capture on the Cl and La targets.  No parity-allowed
asymmetry was seen in those targets.  No asymmetry of
either type was observed with the Cd target.
The results are presented in  Table~\ref{table:results}.
The table includes: values $A_{\rm raw}$, which are the raw
asymmetry data over the analyzed energy range combined by
weighting by number of events; values 
$A_\gamma$ which are obtained from converting the raw asymmetries
to physics asymmetries for each time bin, and then combining
the results weighted by the error on the physics asymmetry in
each bin; and previous results $A_\gamma$~\cite{Vesna,Avenier}.  

The statistical error in the combined parity-violating Cl raw asymmetry 
data is $\pm 2.1 \times 10^{-6}$.  A calculation of the  $\sqrt{N}$ 
statistical error based on the number of neutron capture events
results in an expected error of $\pm 1.9 \times 10^{-6}$.  
The primary sources of uncertainty in this calculation are 
uncertainty in the neutron flux and in the effective solid angle 
of the detectors. The statistical errors on the asymmetries 
follow the expectation of counting statistics given the measured neutron
flux and there is no evidence for extra noise sources.  
Systematic errors are small on the correction factors discussed above:
1\% on $R$, 2\% on $P$, 2\% on $T$, and 5\% on $G$.
Studies which split the data into independent sets to compare asymmetry results
discovered no anomalous effects.

A theoretical estimation of the parity-violating correlation  $ s_n \cdot k_{\gamma} $ 
in polarized cold neutron capture on $^{35}$Cl 
gives $A_\gamma = -(37 \pm 18 ) \times 10^{-6}$~\cite{Bunakov}, 
which is consistent with experiment in both sign and
magnitude. This estimate used the detailed knowledge of
the spectroscopy of  $^{36}$Cl, including the sign cancellations
in the integral asymmetry from the many different transitions in
the  $^{36}$Cl de-excitation spectrum and the gamma-ray energy
weighting appropriate for the current-mode asymmetry measured in
this experiment.

\section{Summary and Prospects}

The parity-violating $ s_n \cdot k_{\gamma} $ neutron capture 
asymmetry measurements  reported here are 
consistent with previous experimental results \cite{Vesna,Avenier}
and of somewhat comparable precision.
However, they were made in a fraction of the run
time (eight hours per target) due to the large flux available with the 
pulsed beamline at LANSCE. In the \npdg\ experiment the parity-violating
asymmetry in $^{35}$Cl will be used
to periodically monitor the performance of the apparatus.

No parity-allowed $ s_n \cdot (k_{\gamma} \times k_n)$
asymmetries were seen in these measurements,
at the level of $7 \times 10^{-6}$.  A large parity-allowed (left-right) asymmetry
would provide a useful technique for alignment of the NPDGamma detector
array by observing the mixing of the known left-right asymmetry
into the up-down detector channels. This alignment is required to suppress 
the size of systematic effects which lead to neutron spin-dependent left-right 
motion of the beam and/or intensity of the gamma-ray angular
distribution. Another possible method
to measure the angle of each detector element to 20~mrad
precision would be to scan a small neutron
capture target in x-y and observe the change in detected
gamma-ray intensity. However, to remove the variable of beam non-uniformity, 
a method using translation of the array (rather than translation
of a capture target) will be used for the full
NPDGamma experiment.

This engineering run, as compared to the full apparatus that 
will be constructed for the NPDGamma experiment, was missing
the following components:  a frame overlap chopper, 
a full-sized $^3$He system, a liquid para-hydrogen target,
a full complement of detectors, and a $^3$He neutron polarization analyzer.
A new beamline at LANSCE for nuclear physics is under construction
with a larger ($9.5 \times 9.5$~cm$^2$ vs.\ $6 \times 6$~cm$^2$)  
$m=3$ supermirror guide.
Scaling of the engineering run errors by run time, proton current, 
moderator brightness, 
guide characteristics of size and transmission,
and detector solid angle, yields an expected 
statistical error on a statistics limited measurement of
$A_\gamma$ for \npdg\ of 
$1.6 \times 10^{-8}$ in 7500 hours 
of delivered neutron beam on FP12 at LANSCE.
All tests show that the experimental design and method 
are sufficient to make a measurement of
this precision for NPDGamma.

\section{Acknowledgments}
The authors would like to thank Mr.\ G.\ Peralta and Dr.\ J.M.\ O'Donnell for 
their technical support during this experiment.
This work was supported in part by the 
U.S.\ Department of Energy
(Office of Energy Research, under Contract W-7405-ENG-36), 
the National Science Foundation (Grant No. PHY-0100348),
and the Natural Sciences and Engineering 
Research Council of Canada.

\vfill

\begin{table}[htbp]
\begin{center}

\renewcommand{\arraystretch}{1.4}

\begin{tabular}{|l|r|r|r|}\hline 
                  
   &  \multicolumn{1}{|c|}{$^{35}$Cl}    
   &  \multicolumn{1}{|c|}{$^{113}$Cd}   
   &  \multicolumn{1}{|c|}{$^{139}$La}    \\ \hline

$A^{\rm PV}_{\rm raw}$
   & $ -8.9 \pm 2.1 $& $ -0.6 \pm 1.5 $& $ -5.2 \pm 2.3 $ \\
$A^{\rm PV}_{\gamma}$
   & $ -29.1 \pm 6.7 $& $ -3.2 \pm 4.5 $& $ -15.5 \pm 7.1 $ \\

\hline

$A^{\rm PV}_{\gamma}$ Leningrad \cite{Vesna} 
  & $ -27.8 \pm 4.9 $& $ -1.3 \pm 1.4$& $ -17.8 \pm 2.2 $ \\
$A^{\rm PV}_{\gamma}$ ILL \cite{Avenier}     
  & $ -21.2 \pm 1.7 $& \multicolumn{1}{|c|}{-} &\multicolumn{1}{|c|}{-} \\ 

\hline

$A^{\rm PA}_{\rm raw}$ 
   & $ -2.0 \pm 2.0 $& $ -2.6 \pm 1.5 $& $ -0.2 \pm 2.2 $ \\
$A^{\rm PA}_{\gamma}$
   & $ -6.5 \pm 6.5 $& $ -7.0 \pm 4.4 $& $ -4.2 \pm 6.7 $ \\

\hline

\multicolumn{4}{|r|}{All asymmetry values are parts per million (units
of $\times 10^{-6}$)}\\

\hline
\end{tabular}

\pagebreak

\renewcommand{\arraystretch}{1.0}

\caption{Results. 
Given errors are statistical only, 
with total systematic errors of approximately 6\%.
parity-violating asymmetries $A^{\rm PV}$ correspond to $ s_n
\cdot k_{\gamma} $,
and parity-allowed asymmetries $A^{\rm PA}$ correspond to 
$s_n \cdot (k_{\gamma} \times k_n)$, where 
$s_n$ is the neutron spin direction, $k_n$ is the neutron momentum vector,
and $k_{\gamma}$ is the photon momentum vector.
}
\label{table:results}
\end{center}     
\end{table}

\pagebreak

\end{document}